\documentclass[aps,prd,preprint,preprintnumbers,nofootinbib,showpacs,floatfix,amsmath,amssymb]{revtex4-1}

\usepackage{graphicx}
\usepackage{subfig}
\usepackage{amsmath}
\usepackage{amssymb}

\newcommand{\ds}{^{}} 
\newcommand{\muX}{\mu_{x}\ds}
\newcommand{\pT}{p_{T}\ds}
\newcommand{\bigO}[1]{\mathcal{O}(#1)}
\newcommand{\MET}{{\not\mathrel{E}}_T\ds}
\newcommand{\abs}[1]{\left | #1 \right |}
\newcommand{\GeV}{\;{\rm GeV}}
\newcommand{\TeV}{\;{\rm TeV}}

\begin{document}

\title{Probing the two Higgs doublet wedge region\\ 
with charged Higgs boson decays to boosted jets}

\author{Keith~Pedersen}
\email{kpeders1@hawk.IIT.edu}
\affiliation{Department of Physics, Illinois Institute of Technology, Chicago, Illinois 60616-3793, USA}

\author{Zack~Sullivan}
\email{Zack.Sullivan@IIT.edu}
\affiliation{Department of Physics, Illinois Institute of Technology, Chicago, Illinois 60616-3793, USA}

\preprint{IIT-CAPP-16-06}

\date{December 11, 2016}


\begin{abstract}
  Two Higgs doublet extensions of the standard model, such as
  supersymmetry, predict the existence of charged Higgs bosons.  We
  explore the reach for TeV-scale charged Higgs bosons through their
  associated production with top quarks, and their decay to boosted
  top jets and $\muX$-tagged boosted bottom jets, at a 14~TeV CERN Large
  Hadron Collider and at a 100~TeV Future Circular Collider.  In
  particular, we show the moderate $\tan\beta$ ``wedge'' region of
  parameter space cannot be probed at the Large Hadron Collider for
  TeV-scale $H^\pm$ because the cross section is too small.  However,
  a 100 TeV future proton collider can close the wedge region below
  2~TeV, and search for $H^\pm$ up to 6~TeV.
\end{abstract}

\maketitle


\section{Introduction}

With the discovery of a 125~GeV boson at the CERN Large Hadron
Collider (LHC)~\cite{Aad:2015zhl}, one which behaves uncannily like
the massive scalar of the standard model's (SM) singular SU(2)
doublet, the question turns to whether an additional scale of physics
can be found in a collider environment.  A generic way to accommodate
another scale of symmetry breaking is to add an additional scalar
field, creating a two Higgs doublet model (2HDM)~\cite{Lee:1973iz}.
2HDMs are commonly associated with supersymmetry (SUSY)\nobreak
~\cite{Fayet:1974pd, Fayet:1976et, Fayet:1977yc, Dimopoulos:1981zb,
  Haber:1984rc}, but they also show up in axion models masking strong
CP violation \cite{Peccei:1977hh, Kim:1986ax} and baryogenesis
\cite{Turok:1990zg, Joyce:1994zt, Funakubo:1993jg}.  2HDM are
primarily characterized by $\tan\beta$ (the ratio of the two vacuum
expectation values) and $\beta-\alpha$ (the doublet mixing angle).
Symmetry breaking produces four scalar Higgs bosons ($h$, $H$,
$H^\pm$) and a pseudo-scalar boson ($A$). If the fine tuning of the
various parameters is minimal, then $h$ is the lightest physical
particle~\cite{Branco:2011iw}.  Given that a wide range of
measurements have effectively ruled out flavor changing neutral
currents at tree-level, realistic 2HDM are restricted to four general
models~\cite{Branco:2011iw}, of which two are worth noting here:
type-I, where all quarks couple to only one of the doublets, and
type-II, where $u_R^i$ and $d_R^i$ couple to opposite doublets (a
requirement of SUSY). We restrict our attention to type-II Higgs
theories.

The SM-like nature of the recently discovered scalar boson (especially
in its per-channel signal strength~\cite{Khachatryan:2016vau}),
constrains many type-II 2HDM rather tightly to the alignment
limit~\cite{Ferreira:2014naa, Craig:2015jba}.  
Here, $\cos(\beta-\alpha)\to0$, forcing $h\to
H^0_{\rm SM}$.\nobreak \footnote{
For $\tan\beta>10$, non-aligned 2HDM are still allowed on a thin trajectory.
} If there is also a near-degeneracy in the masses of $H$, $A$ and $H^\pm$,
a natural consequence of SUSY in the 
decoupling limit~\cite{Gunion:2002zf, Ginzburg:2004vp}, 
then the bosons are kinematically forbidden from decaying to each other.  
This mass degeneracy also occurs in more generic 2HDM models
which favor natural SM alignment without decoupling
(e.g.\ softly broken SO(5)~\cite{Dev:2014yca}).
As such, we explore the degenerate mass sector, 
where the coupling of the heavy charged Higgs boson to the standard model is 
dominated by the heavy third generation.

Detecting $pp\to H/A$ is difficult as both the signal and background
have identical initial and final states ($gg\to q\bar{q}$), and the
resulting interference gives $H/A$ resonances an unusual
shape~\cite{Dicus:1994bm, Asakawa:1999gz} that is easy to mimic with
pure QCD.  Measuring $H/A$ in association with an additional heavy
quark pair eliminates this interference.  For a charged Higgs boson,
associated production is the leading order production mode ($pp\to
H^{\pm}t(b)$), where the associated $b$ can be resummed into the beam
fragments.  This study focuses on the cleaner $tH^\pm+X$ production
channel, as a limit on $H^{\pm}$ is effectively a limit on all four
2HDM scalars.

Assuming quasi-degeneracy of the heavy Higgs masses, one
finds~\cite{Branco:2011iw}
\begin{equation}
\mathcal{L}_{\rm eff}\ds=-H^+ \bar{t} (y_t\ds P_L\ds + 
	y_b\ds P_R\ds)b + {\rm h.c.} \,,
\end{equation}
where $y_t\ds = \sqrt{2}\,m_t\ds\cot\beta/v$ 
and $y_b\ds = \sqrt{2}\,m_b\ds\tan\beta/v$ use running quark masses,
and $P_{L/R}\ds$ are the chiral projection operators.
Appealing to naturalness, while simultaneously keeping 
$y_{tb}\ds=\sqrt{y_t^2+y_b^2}$ perturbative ($y_{tb}\ds\lesssim1$),
leads to the expectation that
\begin{equation}
\tan\beta \ge \frac{\sqrt{2}m_t\ds}{v}
\textrm{~~~~~and~~~~~}
\tan\beta \le \frac{v}{\sqrt{2}m_b\ds} \,,
\end{equation}
which corresponds to $\tan\beta\in\lbrack 0.83, 73. \rbrack$ at
$Q_{\rm 2HDM}=2\TeV$.  At the center of this region
$(\tan\beta=\sqrt{m_t\ds/m_b\ds})$ lies a ``wedge'' of low production
cross section, where the coupling transitions from top-dominated at
low $\tan\beta$ to a bottom-dominated at large $\tan\beta$.  The wedge
obfuscates the investigation of a large swath of interesting parameter
space, as is quite evident in recent experimental searches for $H^\pm$
using 8 TeV LHC data~\cite{Aad:2015typ, Khachatryan:2015qxa}.

The situation should improve in LHC run 2, but the predictions range
from slightly pessimistic for $m_{H^\pm}=0.5$--$1\,{\rm
  TeV}$~\cite{Craig:2015jba} to quite optimistic for
$m_{H^\pm}=0.5$--$2\,{\rm TeV}$~\cite{Hajer:2015gka,
  CEPC-SPPCStudyGroup:2015csa}.  It is our assessment that the
variations in previous estimates are primarily due to choices made
when simulating a standard ``track-vertex'' $b$ tag to suppress QCD
background.  This becomes more difficult as the mass of the charged
Higgs moves above a TeV, as the bottom quark become significantly
boosted, making theoretical predictions sensitive to careful modeling
of real-world tagging efficiencies.

In this work, we predict the experimental reach for $m_{H^\pm}>1\,{\rm
  TeV}$ through its associated production with a top quark, and its
decay to boosted top and boosted bottom jets, in both a generic two
Higgs double model and in SUSY.  In Sec.~\ref{sec:cuts} we describe
our selection cuts and tagging efficiencies in the boosted regime.  In
Sec.~\ref{sec:results} we present our numerical results for the LHC at
14~TeV. We find that the LHC has limited reach to observe a charged
Higgs boson, and so extend our examination to show the reach of a
100~TeV future circular collider (FCC).


\section{Methods}\label{sec:cuts}

Given the disparity between previous predictions for the
reconstruction of $t H^\pm\to t t b$ at large charged Higgs boson
mass, this study concentrates on careful modeling of boosted bottom
jets at the LHC and at a FCC.  In this section, we address
improvements to our existing $\mu_x$ boosted-bottom-jet tag, and
detail improvements to signal selection over previous studies.

\subsection{Bottom-jet tagging}

Track-vertex tags use multivariate information from charged particle
tracks to detect a $b/c$ hadron decay vertex displaced from the
interaction point.  For jet $\pT = \bigO{100\GeV}$, such tags have
high $b$-jet efficiency (${\sim}70\%$) and excellent light jet fake
rates (${\sim}0.1\%$)~\cite{Aad:2015ydr, Chatrchyan:2012jua}, making
them the primary method for $b$ jet tagging at the LHC.

This performance deteriorates as jets approach the TeV regime.  A
highly boosted $b$ jet has tracks which are relatively straight and
very collimated, degrading their individual reconstruction efficiency.
Additionally, the average number of tracks from the bottom hadron
itself is fixed by branching ratios; it does not depend on jet $\pT$.
Thus, as tracking performance degrades for TeV jets, it becomes easier
to miss the limited number of $b$ hadron tracks.  Conversely, the
average number of tracks inside a jet increases with $\pT$, since more
fragmentation produces more particles.  So as light jets ($g$, $u$,
$d$, $s$) become harder, it is easier to find some combination of
tracks which \emph{fake} a displaced vertex~\cite{Aad:2015ydr}.  This
exacerbates the falling signal efficiency with a rising fake rate,
driving $S/B$ even lower. Hence, a common scheme in phenomenological
studies --- treating the nominal $b$-tag efficiency and fake rate as
constant across all $p_T\ds$ --- can lead to over-optimistic
predictions for TeV-scale physics.

These realities spurred the development of the $\muX$
boosted-bottom-jet tag \cite{Pedersen:2015knf}.  The $\muX$ tag is
essentially an angular cut between a muon (from semi-leptonic $b$
hadron decay) and the highly collimated jet ``core'' (the boosted
remnants of the $c$ hadron, along with collinear fragmentation from
the $b$ quark).  The $\mu_x$ tag has been implemented in a public code
for use with fast detector simulators~\cite{Delphes:myDelphes}.  While
the maximum $b$-tagging efficiency of $\muX$ is limited by the overall
branching ratio of semi-muonic $b$-hadron decay (${\sim}19\%$), its
main virtue is that its signal efficiency ($\epsilon_b\ds\approx15\%$)
and fake rate ($\epsilon_{\rm light}\ds\approx0.6\%$) are flat as a
function of jet $\pT$ once boosted kinematics turn on ($p_T$ above
500~GeV).

Previously, we implemented the $\muX$ tag by utilizing the resolution
of the electromagnetic calorimeter and avoided using
tracks~\cite{Pedersen:2015knf}.  In the present study, we improve upon
our prior implementation by allowing $\muX$ to access high-resolution
angular information in tracks to locate the jet core, then calculate
the muon opening angle.  While we find that combining tracking with
normal-resolution calorimetry does not change the tagging efficiency
at 14~TeV, tracking becomes absolutely essential at 100~TeV.  The
large radius~(6~m) and strong magnetic field~(6~T) of the hypothetical
FCC tracking system~\cite{deFavereau:2013fsa} smears the charged
constituents in~$\phi$, reducing the correlation between charged
tracks and multi-TeV tower jets.

\subsection{Signal selection}

There are two major production modes for $tH^+$ at a proton collider:
the ``$4b$'' final state $g g \to \lbrack H^{+} \to \bar{t} b\rbrack t
\bar{b}$ (with $t\to b W^+$), and the ``$3b$'' final state $g b \to
(H^{+} \to \bar{t} b) t$.  Since the $3b$ final state is the dominant
mode, accounting for at least 60\% of the total cross section for all
masses, the inclusive ($3b$ + $4b$) final state is a natural starting
point.  This requires tagging a boosted bottom jet and two tops: a
boosted top jet from the $H^{\pm}$ decay, and a much softer,
resolvable, associated top.

Using the $\muX$ tag to identify the boosted-$b$ jet unavoidably
selects events containing hard neutrinos from semi-leptonic $B$ hadron
decay.  This smears the missing energy of any leptonically decaying
tops, reducing the effectiveness of $\MET$ for top identification or
reconstruction, and limiting $H^{\pm}$ mass resolution if the boosted
top decays leptonically.  These limitations are easily side-stepped by
using only the fully hadronic decay of the boosted top, tagging the
unique shape of $t\to W^{+} b$ merged into a single ``fat''
jet~\cite{CMS:2009lxa}.  Conversely, the associated top is slow enough
to be resolved into isolated daughters, so its fully hadronic final
state is quite susceptible to QCD background.  It is safer to resolve
the associated top into an isolated lepton ($e/\mu$) and a $b$ jet
(which is soft enough that high-efficiency track tags remain robust).

The $t\bar{t}$ portion of the inclusive final state provides multiple
handles to suppress pure multijet background, leaving $ttj+X$ the
dominant background (where $j=guds$).  Here, the light flavored jet is
both hard and ``mis-tagged'' as a primary boosted-$b$ jet.  This
usually occurs when the jet showers $g\to b\bar{b}$, creating a real
$B$-hadron inside a jet of light-flavor origin.  The sub-dominant
background is $tt(bb/cc)$ --- effectively the same final state, but
with the gluon splitting at a much higher scale.  Other final states
(e.g.\ $tjj+X$ and $ttbj$) are found to be negligible.

Event reconstruction begins with jet reconstruction.  First,
``narrow'' jets are clustered using an anti-$kt$ algorithm with
$R=0.4$ \cite{Cacciari:2008gp}, and ``fat'' jets are clustered using a
Cambridge-Aachen algorithm with $R=0.8$ \cite{Dokshitzer:1997in}.
Both boosted jets must have $p_{T j}\ds\ge350\GeV$, and all jets must
have $p_{T j}\ds\ge20(40)\GeV$ for 14(100) TeV collisions.
Additionally, all jets must have $\abs{\eta_j\ds}<2.1(3.0)$, so that
the edge of the tracker lies outside the clustering radius of narrow
jets.  We require exactly one isolated lepton with $p_T^{\rm
  lepton}>15(25)$~GeV.  The lepton is considered isolated if
$p_{T}^{\rm lepton}/\sum_{i}\ds p_{T}^{i} < 5\%$ for all tracks and
towers within a cone of $\Delta R < 10\GeV / p_T^{\rm lepton}$, as
prescribed in a recent experimental search~\cite{Aad:2013nca}.
Additionally, the lepton cannot fall within a $\Delta R=R_{\rm
  cluster}$ cone surrounding any of the candidate jets.

Narrow jets are sorted by $\pT$ (high to low), and the first narrow
jet which is $\muX$ tagged becomes the boosted $b$ candidate.  To
exclude the situation where the boosted top decays leptonically (and
the associated top hadronically), we require that the boosted-$b$ plus
lepton system has a mass inconsistent with a top quark
($m_{bl}>172\GeV$).  This cut is primarily used to properly model the
$ttj+X$ background, but is redundant in other systems because it
effectively overlaps the requirement that the lepton reside outside of
the boosted-$b$ jet.

Next, fat jets are sorted by $\pT$, and the first one which has a
boosted hadronic top tag is the boosted top candidate.  We then
require that $\Delta R_{bt}\ge2$ and $\abs{\Delta \eta_{bt}} \le 2$
for the two boosted candidates.  The latter cut is used to restrict
$t$-channel background from hardening the tail of the $m_{bt}$
distribution, although it removes about a fifth of all $H^{\pm}$
(whose isotropic decay is minimally boosted in the transverse
direction, due to its large mass).  We do not impose any constraints
on the mass of the boosted top jet, as these are already built into
the boosted top tag efficiency.

We then attempt to reconstruct the associated top by finding a $b$ jet
compatible with the isolated lepton.  From the set of narrow jets
whose $\pT$ is smaller than the boosted $b$, we take at most two jets
which are $b$-tagged and reside outside an $R=1.2$ cone around
the boosted top (which should contain its own $b$ jet).  We then
attempt to find a $b$-lepton system with $\pT$ less than the boosted
top, and an invariant mass consistent with a top quark missing its
neutrino ($70\GeV<m_{bl}<180\GeV$, where the slightly elevated ceiling
permits detector smearing).  If two $b$ candidates pass these cuts,
the one whose $m_{bl}$ is above 110~GeV is selected; if both are above
110~GeV, the one which is closer to 110~GeV is selected.

The total branching ratio of the hadronic/leptonic $t\bar{t}$ decay
(14\%), combined with the efficiency of the two boosted flavor tags
($\epsilon_b\ds\approx0.14$ and $\epsilon_t\ds\approx0.45$) and the
event shape cuts for the inclusive final state, produce an overall
$H^{\pm}$ acceptance of $\bigO{0.1\%}$.  The QCD background acceptance
is an order of magnitude lower, though a more important consideration
is the ratio of $ttj + X$ to $tt(bb/cc)$.  For the inclusive cuts, the
ratio is consistently about 5:1, which is small enough that there is
no clear benefit to independently reconstructing the $4b$ final state,
as was previously done \cite{Hajer:2015gka,Craig:2015jba}, since the
process is already signal constrained at the LHC.

\section{Results}\label{sec:results}

We calculate all cross sections using a generic 2HDM from
FeynRules~\cite{Christensen:2008py, Alloul:2013bka, Degrande:2014vpa,
  Duhr:2HDM} with {\sc MadGraph}~5~v2.3.3~\cite{Alwall:2014hca} and
the CT14llo parton distribution functions~\cite{Dulat:2015mca}.
Events are showered and hadronized using {\sc Pythia} 8.210
\cite{Sjostrand:2007gs, Sjostrand:2006za}, and reconstructed using
\texttt{FastJet}~3.1.3~\cite{Cacciari:2011ma} and the {\sc
  Delphes}~3~\cite{deFavereau:2013fsa} detector simulation.  For the
14~TeV analysis, we modify the ATLAS card supplied with {\sc Delphes}
to simulate the $\muX$ boosted $b$ tag (using the
\texttt{MuXBoostedBTag} module available on
GitHub~\cite{Delphes:myDelphes}).  Both the track-based $b$ tag and
the boosted top tag are applied using a functional form of the tagging
efficiency based upon jet $\pT$.  For the track-vertex $b$ tag, we use
the run~2 efficiency from the ATLAS card (based upon
Ref.~\cite{ATLAS:run-2-bTag}), and for the top tag, we use the
efficiencies depicted in Ref.~\cite{CMS:2009lxa}, which closely match
those given in more recent publications~\cite{Aad:2016pux,
  CMS:2014fya}.  At 100~TeV, we use the FCC card supplied with {\sc
  Delphes} (again modified to simulated $\muX$), with two major
changes: (i)~we use the same track-vertex $b$ tagging efficiency
formula used for 14~TeV and (ii)~we use a more conservative tracking
domain ($\abs{\eta}\le3.5$).

At both collider energies, we use {\sc Delphes}' ``EFlow'' objects
(which subtracts track energy from the calorimeter towers they strike,
after both tracks and towers have their energy smeared).  We then
cluster jets from tracks (minus isolated leptons) and track-subtracted
towers.  To estimate the neutrino $\MET$ inherent to the $\muX$ tag,
we simply double the momentum of the tagging
muon~\cite{Pedersen:2015knf}.  This does a reasonably effective job of
reconstructing the $H^{\pm}$ peak, allowing us to use a mass window of
$[0.9, 1.15]\times M_{H^{\pm}}$ at both 14 and 100~TeV to capture the
majority of the signal.  Without neutrino estimation, the $H^{\pm}$
peak has a noticeably longer low-mass tail.

\subsection{$tH^\pm \to ttb$ in a generic 2HDM}

\begin{figure*}[htb]
\subfloat[\label{fig:ytb-14}]{
\includegraphics[width=0.5\textwidth]{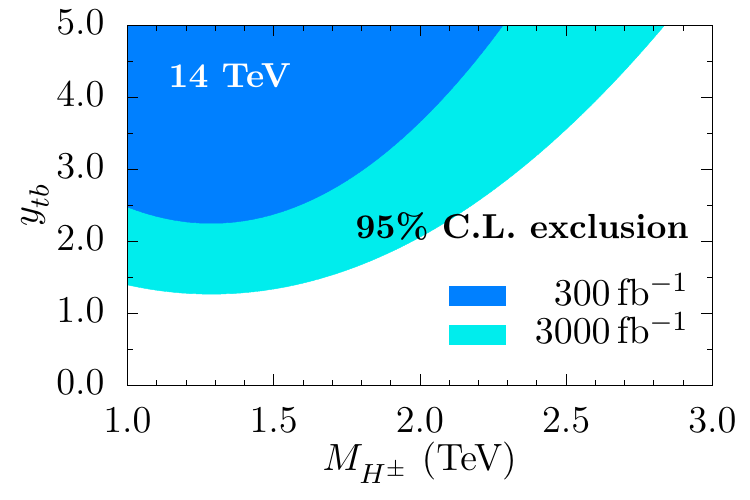}}
\subfloat[\label{fig:tanBeta-14}]{
\includegraphics[width=0.5\textwidth]{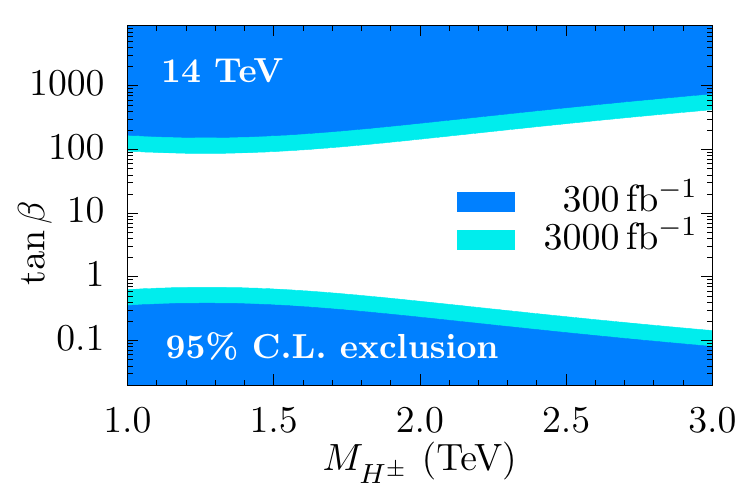}}
\caption{Predicted exclusion regime, at a 95\%~confidence level, for a
  generic 2HDM at a 14~TeV LHC in terms of (a) the effective Yukawa
coupling $y_{tb}\ds$, and (b) the corresponding $\tan\beta$.}
\label{fig:exclusion-14}
\end{figure*}

We first explore the reach for a charged Higgs boson produced in
association with a top quark for a generic 2HDM.  We convert the
leading order $y_{tb}\ds$ used by MadGraph to a next-to-leading
order $y_{tb}\ds$ by using the running quark masses at one-loop in QCD
\cite{Carena:1999py}, which shifts the center of the $\tan\beta$ wedge
upwards.  In Fig.~\ref{fig:exclusion-14}, we show the 95\% confidence
level (C.L.)  limit for $H^{\pm}$ exclusion at a 14~TeV LHC with 300
or 3000~${\rm fb^{-1}}$ of data.  In order to compare directly with
Refs.~\cite{Craig:2015jba,Hajer:2015gka}, we show (a) the limit
obtainable on the effective Yukawa coupling $y_{tb}$, and (b) the
corresponding region of $\tan\beta$ probed.  It turns out that the
only Yukawa couplings $y_{tb}$ or values of $\tan\beta$ that can be
probed at the LHC are on the border of the non-perturbative regions of
parameter space.

The accessible region of parameter space at the LHC is entirely
limited by the production cross section, as $S/B=\bigO{1/2}$ across
the entire mass range.  Because the $tH^{\pm}$ cross section at 14~TeV
is quite small, the reach in $\tan\beta$ is poor at the LHC.  Once
$M_{H^{\pm}}$ surpasses 2~TeV, the $H^{\pm}$ begins to grow noticeably
off-shell, which weakens the narrow width approximation we use to
extrapolate from our working value of $\tan\beta$ to the 95\% limit.
The loss of reach approaching 1~TeV is due to signal/background
attenuation; a combination of the 350~GeV minimum $\pT$ cut imposed on
both boosted jets and the swiftly diminishing efficiency of both
boosted flavor tags below 500~GeV.  Given the presence of this
feature, our results are consistent with extending the predictions of
Ref.~\cite{Craig:2015jba} into the TeV regime.  Charged Higgs bosons
are unlikely to be observed at the LHC.

\begin{figure*}[htb]
\subfloat[\label{fig:ytb-100}]{
\includegraphics[width=0.5\textwidth]{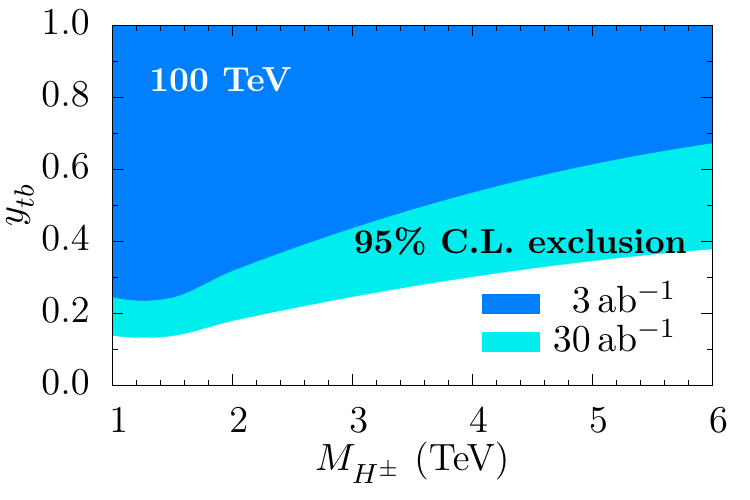}}
\subfloat[\label{fig:tanBeta-100}]{
\includegraphics[width=0.5\textwidth]{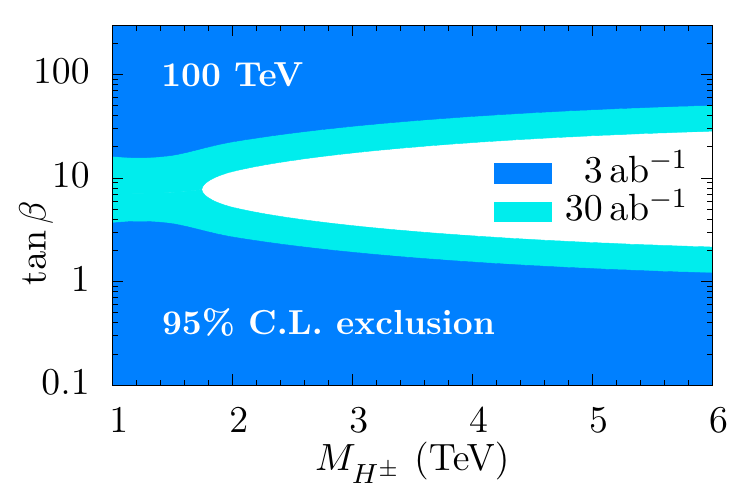}}
\caption{Predicted exclusion regime, at a 95\%~confidence level, for
  a generic 2HDM at a 100~TeV FCC in terms of (a) the effective Yukawa
  coupling $y_{tb}\ds$, and (b) the corresponding $\tan\beta$.}
\label{fig:exclusion-100}
\end{figure*}

The $tH^\pm$ cross section is strongly dependent on collider energy.
A 100 TeV collider, such as a FCC, promises significantly more reach
for charged Higgs bosons.  At 100~TeV, the reach becomes background
limited, with $S/B$ rising from ${\sim}1\%$ at 1~TeV to ${\sim}5\%$ at
6~TeV.  In Fig.~\ref{fig:exclusion-100} we observed that the reach in
effective Yukawa coupling is an order-of-magnitude better than at the
LHC.  This allows the wedge region to close as the integrated
luminosity rises above 3~${\rm ab^{-1}}$ up to a charged Higgs mass of
2~TeV.  While this analysis is robust, more sophisticated techniques
--- boosted decision trees (BDT) or neural nets (NN) --- might improve
the reach.  However, since BDT/NN techniques are highly dependent on
the quality of the observables with which they train, it is difficult
to make accurate predictions this far from a realized 100~TeV detector
system, especially using a fast detector simulator.  Regardless, our
results suggest that search for TeV-scale charged Higgs bosons is the
domain of future colliders.

\subsection{$tH^\pm \to ttb$ in a supersymmetric model}

One-loop corrections in the minimal supersymmetric standard model
(MSSM) modify the fermionic couplings to $H^{\pm}$ bosons.  The effect is
most significant for the bottom quark~\cite{Dawson:2007wh,Carena:1999py},
and can be absorbed into the Yukawa coupling as
\begin{equation}
y_b^{\rm SQCD} = y_b \frac{1}{1+\Delta m_b\ds} \,,
\end{equation}
(here we ignore supersymmetric electroweak corrections, using only
those from supersymmetric QCD).  $\Delta m_b\ds$ explicitly depends on
the the gluino mass, the mass of the two bottom squark eigenstates and
$\mu$, the mass parameter coefficient of the $\epsilon_{ij}\ds H^1_i
H^2_j$ term in the superpotential.  In the quasi-degenerate limit,
where all these mass parameters are of equal size, only the sign of
$\mu$ survives~\cite{Carena:1999py}.  At large $\tan\beta$
($\sin\beta\approx1)$
\begin{equation}
\Delta m_b \approx {\rm sign}(\mu) \frac{\alpha_s\ds(Q_{\rm SUSY}\ds)}{3 \pi}\tan\beta \,,
\end{equation}
where $Q_{\rm SUSY}\ds$ is the heavy SUSY scale (which we take to be
10~TeV, although the result is not heavily dependent upon the choice
of $Q_{\rm SUSY}\ds$, since $\alpha_s$ runs slowly above a few TeV).

\begin{figure*}[htb]
\subfloat[\label{fig:tanBeta-14-SUSY+}]{
\includegraphics[width=0.5\textwidth]{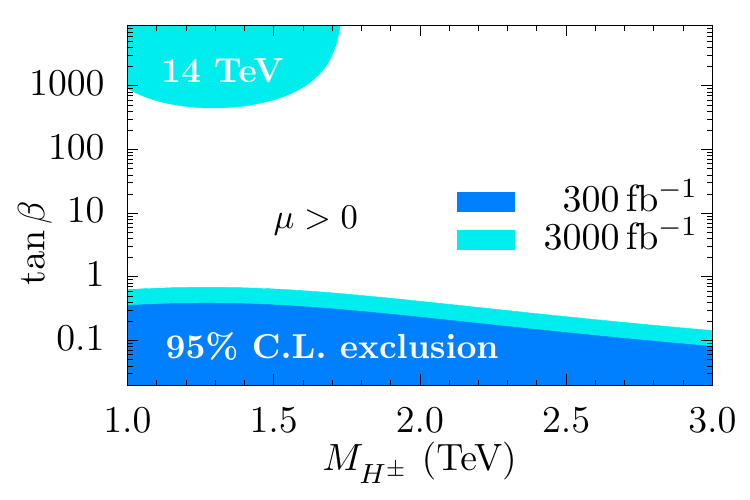}}
\subfloat[\label{fig:tanBeta-14-SUSY-}]{
\includegraphics[width=0.5\textwidth]{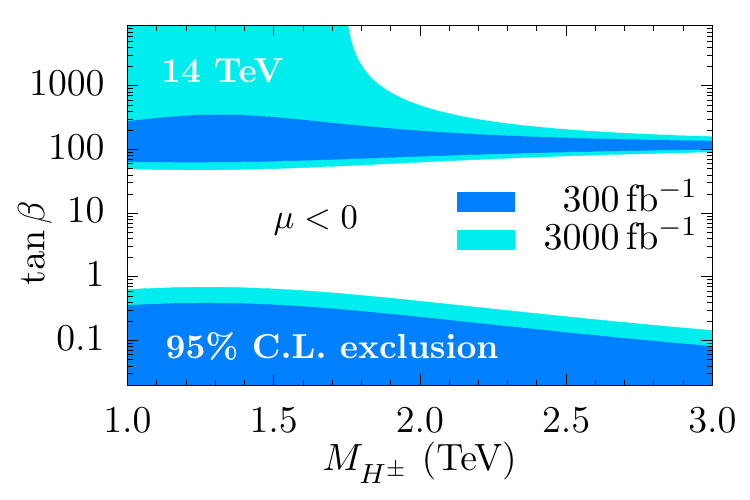}}
\caption{Predicted exclusion regime, at a 95\%~confidence level, for
  the MSSM at a 14~TeV LHC, taking the sign of $\mu$ to be (a)
  positive, or (b) negative.}
\label{fig:exclusion-SUSY-14}
\end{figure*}

Comparing Fig.~\ref{fig:exclusion-SUSY-14} to
Fig.~\ref{fig:exclusion-14}, it is readily apparent that the $\Delta
m_b\ds$ correction has a significant impact on the reach at 14~TeV,
where the production cross section is so small that only very large
$\tan\beta$ are accessible.  For a positive $\mu$, the $\Delta m_b\ds$
correction counteracts the cross section enhancement of large
$\tan\beta$, shifting high $\tan\beta$ parameter space completely out
of reach.  Conversely, the negative $\mu$ correction enhances the
cross section beyond the generic 2HDM in a small region of
$\tan\beta\sim 100$, but decreases it at larger values of $\tan\beta$.
At small values of $\tan\beta < 0.5$, the top-quark Yukawa becomes so
large the theory is non-perturbative.  If charged Higgs boson searches
are difficult at the LHC in a generic 2HDM, in SUSY they are nearly
impossible.

\begin{figure*}[htb]
\subfloat[\label{fig:tanBeta-100-SUSY+}]{
\includegraphics[width=0.5\textwidth]{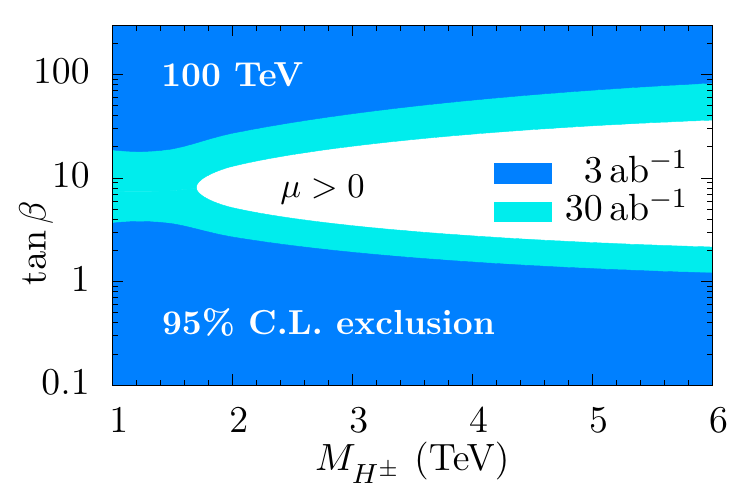}}
\subfloat[\label{fig:tanBeta-100-SUSY-}]{
\includegraphics[width=0.5\textwidth]{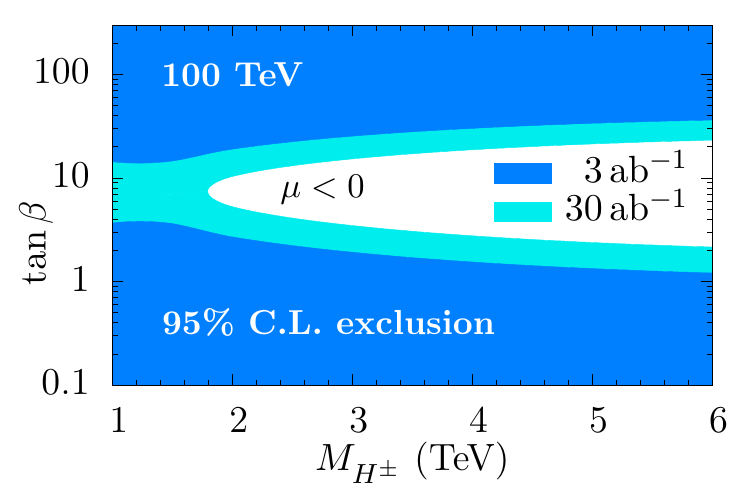}}
\caption{Predicted exclusion regime, at a 95\%~confidence level, for
  the MSSM at a 100~TeV FCC, taking the sign of $\mu$ to be (a)
  positive, or (b) negative.}
\label{fig:exclusion-SUSY-100}
\end{figure*}

In stark contrast, Fig.~\ref{fig:exclusion-SUSY-100} shows that the
effect of $\Delta m_b\ds$ is noticeable at a 100~TeV collider, but it
manifests only as a moderate shift in the upper bound of the wedge,
without a dramatic change in shape. This serves to underline the
nature of the $\Delta m_b\ds$ effect; for a \emph{signal limited}
search (14~TeV), it is very important, while for a \emph{background
  limited} search (100~TeV) it is more-or-less negligible.  The lack
of sensitivity to SUSY corrections at 100~TeV demonstrates the low
model dependence in the reach for charged Higgs bosons at a future
collider.

\section{Conclusions}

We examine the predicted experimental reach for charged Higgs bosons
in $tH^\pm\to ttb$ at both the LHC and at a 100~TeV future collider,
using a type-II two Higgs doublet model with mass degenerate heavy
Higgs bosons.  In the limit where $H^\pm$ couples mostly to $tb$, we
find that the LHC has access only to relatively large effective Yukawa
couplings $y_{tb}\ds$ when $m_{H^\pm}>1\,\TeV$ --- confirming and
extending the expectations from Ref.~\cite{Craig:2015jba}.
Additionally, we find that supersymmetric corrections to the bottom
Yukawa coupling are large, and further reduce sensitivity to a MSSM
charged Higgs boson at the LHC.  These findings indicate that a
next-generation collider will probably be necessary to examine
TeV-scale charged Higgs bosons that couple strongly to the third
generation of quarks.  In comparison to more optimistic
predictions~\cite{Hajer:2015gka}, we stress the importance of using
realistic $b$-tagging efficiencies
\cite{Pedersen:2015knf,Delphes:myDelphes} in phenomenological
predictions covering TeV-scale physics.

Our particular choice of 2HDM (type-II with degenerate masses) ensures
that $H^\pm tb$ is the only pertinent coupling.  A less restrictive
model (e.g.\ where $H^\pm$ couples to charm
\cite{Altmannshofer:2016zrn}), or one with alternate decay channels,
such as $H^\pm\to W^\pm H$, may still be visible at the LHC given
sufficient integrated luminosity.  In those cases, one can convert our
limit on $y_{tb}\ds$ to a limit on cross-section times branching
fraction for the channel $tH^\pm\to ttb$ in those models.

Finally, we find a 100~TeV proton collider has the potential to close the
moderate $\tan\beta$ ``wedge'' region below 2~TeV.  While the charged
Higgs-top associated channel will be background limited at such a
machine, charged Higgs bosons with masses up to 6~TeV can be probed
with very little dependence on model parameters (such as the sign of
the $\mu$-parameter in SUSY).  Hence, a future circular collider shows
great promise in shedding light on the structure of multiplets in the
Higgs boson sector.

\begin{acknowledgments}
  We thank Carlos Wagner for his help with the SUSY $\Delta m_b\ds$
  correction.  This work was supported by the U.S.\ Department of
  Energy under award No.\ DE-SC0008347.
\end{acknowledgments}

\end{document}